\documentclass[twocolumn]{aastex}
\usepackage{graphicx,floatrow,amsmath,gensymb,rotating}
\usepackage{subfigure}
\usepackage{natbib}
\usepackage{xcolor}
\usepackage{threeparttable,booktabs}
\usepackage{etoolbox}

\def\pdegree{^{\circ}\mskip -7.6mu.\,}

\newcommand{\rvs}[1]{{#1}}
\def\etal{{et~al.\null}}

\newcommand{\oiii}{\hbox{[O\,{\scriptsize III}]}}

\newcommand{\oii}{\hbox{[O\,{\scriptsize II}]}}

\begin{document}

\title{The Pre-explosion Environments and The Progenitor of SN~2023ixf from the Hobby Eberly Telescope Dark Energy Experiment (HETDEX)}

\author[0000-0001-5561-2010]{Chenxu Liu}
\affiliation{South-Western Institute for Astronomy Research, Yunnan University, Kunming, Yunnan, 650500, People's Republic of China}
\email{cxliu@ynu.edu.cn}

\author[0009-0000-4068-1320]{Xinlei Chen}
\affiliation{South-Western Institute for Astronomy Research, Yunnan University, Kunming, Yunnan, 650500, People's Republic of China}

\author[0000-0002-8700-3671]{Xinzhong Er}
\affiliation{South-Western Institute for Astronomy Research, Yunnan University, Kunming, Yunnan, 650500, People's Republic of China}

\author[0000-0003-2307-0629]{Gregory R. Zeimann}
\affiliation{McDonald Observatory, The University of Texas at Austin, 2515 Speedway Boulevard, Austin, TX 78712, USA}

\author[0000-0001-8764-7832]{J{\'o}zsef Vink{\'o}}
\affiliation{Department of Astronomy, The University of Texas at Austin, 2515 Speedway Boulevard, Austin, TX 78712, USA}
\affiliation{ Konkoly Observatory,  CSFK, Konkoly-Thege M. \'ut 15-17, Budapest, 1121, Hungary}
\affiliation{ELTE E\"otv\"os Lor\'and University, Institute of Physics, P\'azm\'any P\'eter s\'et\'any 1/A, Budapest, 1117 Hungary}
\affiliation{Department of Experimental Physics, University of Szeged, D\'om t\'er 9, Szeged, 6720, Hungary}

\author[0000-0003-1349-6538]{J.\ Craig Wheeler}
\affiliation{Department of Astronomy, The University of Texas at Austin, 2515 Speedway Boulevard, Austin, TX 78712, USA}

\author[0000-0002-2307-0146]{Erin Mentuch Cooper}
\affiliation{Department of Astronomy, The University of Texas at Austin, 2515 Speedway Boulevard, Austin, TX 78712, USA}
\affiliation{McDonald Observatory, The University of Texas at Austin, 2515 Speedway Boulevard, Austin, TX 78712, USA}

\author[0000-0002-8925-9769]{Dustin Davis}
\affiliation{Department of Astronomy, The University of Texas at Austin, 2515 Speedway Boulevard, Austin, TX 78712, USA}

\author[0000-0003-2575-0652]{Daniel J. Farrow}
\affiliation{Centre of Excellence for Data Science, Artificial Intelligence and Modelling,
University of Hull, Cottingham Road, Hull, HU6 7RX, UK}
\affiliation{E.A. Milne Centre for Astrophysics, University of Hull, Cottingham Road, Hull, HU6 7RX, UK}

\author[0000-0002-8433-8185]{Karl Gebhardt}
\affiliation{Department of Astronomy, The University of Texas at Austin, 2515 Speedway Boulevard, Austin, TX 78712, USA}

\author[0000-0001-5737-6445]{Helong Guo}
\affiliation{South-Western Institute for Astronomy Research, Yunnan University, Kunming, Yunnan, 650500, People's Republic of China}

\author[0000-0001-6717-7685]{Gary J. Hill}
\affiliation{McDonald Observatory, The University of Texas at Austin, 2515 Speedway Boulevard, Austin, TX 78712, USA}
\affiliation{Department of Astronomy, The University of Texas at Austin, 2515 Speedway Boulevard, Austin, TX 78712, USA}

\author[0000-0002-1496-6514]{Lindsay House}
\affiliation{Department of Astronomy, The University of Texas at Austin, 2515 Speedway Boulevard, Austin, TX 78712, USA}

\author[0000-0002-0417-1494]{Wolfram Kollatschny}
\affiliation{Institut f\"ur Astrophysik und Geophysik, Universit\"at G\"ottingen, Friedrich-Hund Platz 1, 37077 G\"ottingen, Germany}

\author[0009-0009-9343-090X]{Fanchuan Kong}
\affiliation{South-Western Institute for Astronomy Research, Yunnan University, Kunming, Yunnan, 650500, People's Republic of China}

\author[0000-0001-7225-2475]{Brajesh Kumar}
\affiliation{South-Western Institute for Astronomy Research, Yunnan University, Kunming, Yunnan, 650500, People's Republic of China}

\author[0000-0003-0394-1298]{Xiangkun Liu}
\affiliation{South-Western Institute for Astronomy Research, Yunnan University, Kunming, Yunnan, 650500, People's Republic of China}

\author[0000-0002-7327-565X]{Sarah Tuttle}
\affiliation{Department of Astronomy, University of Washington, Seattle, Physics \& Astronomy Building, Seattle, WA, 98195, USA}

\author[0000-0002-7714-6310]{Michael Endl}
\affiliation{Department of Astronomy, The University of Texas at Austin, 2515 Speedway Boulevard, Austin, TX 78712, USA}

\author{Parker Duke}
\affiliation{Department of Astronomy, The University of Texas at Austin, 2515 Speedway Boulevard, Austin, TX 78712, USA}

\author[0000-0001-9662-3496]{William D.\ Cochran}
\affiliation{McDonald Observatory, The University of Texas at Austin, 2515 Speedway Boulevard, Austin, TX 78712, USA}
\affiliation{Center for Planetary Systems Habitability, The University of Texas at Austin, Austin, TX 78712, USA}

\author[0000-0002-2510-6931]{Jinghua Zhang}
\affiliation{South-Western Institute for Astronomy Research, Yunnan University, Kunming, Yunnan, 650500, People's Republic of China}

\author{Xiaowei Liu}
\affiliation{South-Western Institute for Astronomy Research, Yunnan University, Kunming, Yunnan, 650500, People's Republic of China}

\shorttitle{Pre-explosion Environments of SN~2023ixf}
\shortauthors{Liu et al.}

\begin{abstract}

Supernova (SN) 2023ixf was discovered on May 19th, 2023. The host galaxy, M101, was observed by the Hobby Eberly Telescope Dark Energy Experiment (HETDEX) collaboration over the period April 30, 2020 -- July 10, 2020, using the Visible Integral-field Replicable Unit Spectrograph (VIRUS; $3470\lesssim\lambda\lesssim5540$~\AA) on the 10-m Hobby-Eberly Telescope (HET). The fiber filling factor within $\pm30\arcsec$ of SN~2023ixf is 80\% with a spatial resolution of 1\arcsec. The $r<5.5\arcsec$ surroundings are 100\% covered. This allows us to analyze the spatially resolved pre-explosion local environments of SN~2023ixf with nebular emission lines. The 2-dimensional (2D) maps of the extinction and \rvs{the star-formation rate (SFR) surface density ($\Sigma_{\rm SFR}$)} show weak increasing trends in the radial distributions within the $r<5.5\arcsec$ regions, suggesting lower values of extinction and SFR in the vicinity of the progenitor of SN~2023ixf. The median extinction and that of the surface density of SFR within $r<3\arcsec$ are $E(B-V)=0.06\pm0.14$, and $\Sigma_{\rm SFR}=10^{-5.44\pm0.66}~\rm M_{\sun}\cdot yr^{-1}\cdot arcsec^{-2}$. There is no significant change in extinction before and after the explosion.
The gas metallicity does not change significantly with the separation from SN~2023ixf. The metal-rich branch of the $R_{23}$ calculations indicates that the gas metallicity around SN~2023ixf is similar to the solar metallicity ($\sim Z_{\sun}$). The archival deep images from the Canada-France-Hawaii Telescope Legacy Survey (CFHTLS) show a clear detection of the progenitor of SN~2023ixf in the $z$-band at $22.778\pm0.063$ mag, but non-detections in the remaining four bands of CFHTLS ($u,g,r,i$). The results suggest a massive progenitor of $\approx$ 22~$\rm M_\sun$.

\end{abstract}

\keywords{supernovae: individual: 2023ixf, galaxy: individual: M101}

\section{Introduction}
\label{sec_intro}
The life of a massive star ends with a dramatic and energetic explosion known as a core-collapse supernova \citep[CCSN;][]{snex}. These impressive events present a unique opportunity to delve into the physics of massive star evolution and the creation of heavy elements through nucleosynthesis. It has been discovered that the progenitor stars of CCSNe experienced eruptive mass loss in the years prior to the core-collapse \citep[e.g.][]{2014ARA&A..52..487S}. Pre-SN outbursts are often observed especially in Type IIn that display narrow emission lines for extended times after the explosion \citep[e.g.][]{2014ApJ...789..104O} as a consequence of the shock interaction between the slow-moving circumstellar medium (CSM) and the ejected material of the explosion \citep{2014ARA&A..52..487S}. However, the underlying physical mechanism that triggers those outbursts and the mass loss rates involved remain elusive \citep[e.g.][]{
2020MNRAS.492.5994B,2020AJ....160..145H,2020MNRAS.494.5230G,2023ApJ...944...34O,2022ApJ...935...31H}. Direct identification of progenitors and the environments of stars via pre-explosion imaging becomes crucial in understanding these unsolved questions. Nonetheless, resolving individual stars in external galaxies at large distances is challenging, making the detection and study of SN events in nearby galaxies particularly important.

SN~2023ixf was reported on 2023 May 19.727 UT in M101 \citep[\rvs{NGC~5457},][]{2023TNSTR1158....1I}, and was classified as a Type II SN \citep{2023TNSAN.119....1P}. Being the SN event of the smallest distance in the past decade, intensive archival observations are available to identify the progenitor and to constrain the pre-explosion environment \citep[e.g.][]{Pledger2023,Jencson2023,JG23,2023arXiv230604722K,2023arXiv230606162N,2023arXiv230610783S,2023arXiv230615270S,Hira23,2023arXiv230702539D, VanDyk2023}.

M101 is an Scd-type galaxy with active star formation \citep{Lin2013}. Several SNe events have been identified in the past century, e.g. the extensively studied SN 2011fe. Detailed comparison of the environments before and after the SN explosion may provide a direct constraint to the nature of SN feedback in the study of galaxy formation. 
Here we report the detection of the candidate progenitor star in the archival data from the Canada-France-Hawaii Telescope Legacy Survey (CFHTLS). We analyze the pre-explosion environments of the progenitor using integral field spectroscopic observation from the Hobby Eberly Telescope Dark Energy Experiment (HETDEX).
In this work, we use the distance estimate of the host galaxy $6.85\pm0.15$ Mpc reported by \citet{2022ApJ...934L...7R} and the redshift estimate of  $z=0.000804$ \citep{2023TNSAN.119....1P}. The coordinates of SN~2023ixf are (RA$=210.910674637$, Dec$=+54.3116510708$, J2000)\footnote{\url{https://www.wis-tns.org/object/2023ixf}}.

\section{Data}
\label{sec_data}

\subsection{HETDEX}
\label{sec_hetdex}

HETDEX \citep{Gebhardt2021} is an untargeted spectroscopic survey aimed at constraining the nature of dark energy by mapping out the three-dimension distribution of  Ly$\alpha$ emitting galaxies between the redshifts  $1.9 < z < 3.5$.  This survey is being carried out on the upgraded 10-m Hobby-Eberly Telescope (HET) with a segmented spherical primary mirror \citep{Ramsey1994,Hill2021} and a 22\arcmin\ aberration-corrected field of view. The upgraded HET includes a new instrument called the Visible Integral-field Replicable Unit Spectrograph, which is a massively replicated, fiber-fed integral field spectrograph \citep{Hill2021} designed specifically to identify emission-line objects. VIRUS consists of 78 Integral-Field Units (IFUs), each containing 448 $1 \farcs 5$-diameter fibers packed into a $51\arcsec \times 51\arcsec$ array.  Each IFU feeds two low-resolution ($R \sim 800$) spectrographs, together covering the wavelength range $3470 \lesssim \lambda \lesssim 5540$~\AA\null. At the project's nominal depth, the spectra reach a $5\sigma$ completeness of $3.5 \times 10^{-17}$~ergs~cm$^{-2}$~s$^{-1}$ at [\ion{O}{3}] $\lambda5007$.

A typical HETDEX observation uses a 3-point dither pattern to fill in the gaps between fibers, with each dithered exposure being 6 minutes in length. The IFUs themselves are distributed in a grid, with each IFU separated from its nearest neighbor by $50\arcsec$ (except in the very center of the array, where other HET instruments are located).  The result is that each HETDEX observation has a fiber-fill factor of 1 in 4.5.  M101 lies within the footprint of the HETDEX survey, but to facilitate studies of the galaxy, additional pointings were defined to give a near complete fill factor of 1. The data of the full M101 observations were collected over the period April 30, 2020 -- July 10, 2020, during a time when $\sim 70$ of the planned 78 VIRUS IFUs were operational. 

In total, there are 21 HETDEX observations covering the M101 field. Unlike the main HETDEX survey design, the footprint was chosen to provide as complete sky coverage as possible with tiling designed to fill in the gaps between the IFU array of the VIRUS spectrograph. We generate a datacube mosaic from these set of observations using a different reduction method than that described in \cite{Gebhardt2021}. The HETDEX observations of M101 were processed by \texttt{Remedy}\footnote{\url{https://github.com/grzeimann/Remedy}}. Here we briefly  summarize the data reduction and calibration of the M101 observations but further details are found in Zeimann \etal\ (in preparation). Astrometry is determined by matching wavelength collapsed images produced from the VIRUS spectra with stars within the Pan-STARRS \citep{Chambers2016} Data Release 2 catalog. The root mean-square (rms) of the stars positions about the solution is $\lesssim 0.25\arcsec$\ and the rms of the solution is typically less than $\lesssim 0.05\arcsec$. 

In typical HETDEX observations, as described in \cite{Gebhardt2021}, flux calibration is determined using field stars of the Sloan Digital Sky Survey \citep[SDSS;][]{York2000}. However, in the case of M101, where there are no SDSS fields available and we simply adopt an average throughput curve from  all HETDEX observations, adjusted for the conditions of that observation.  We normalize our response curves to the monochromatic fluxes obtained from the narrow-band [\ion{O}{3}] $\lambda5007$ images of  Herrmann \etal\ (in preparation). To do this, we apply the narrow-band filter's transmission curve\footnote{\url{https://www.noao.edu/kpno/mosaic/filters/k1014.html}} to VIRUS spectra to create a [\ion{O}{3}] $\lambda5007$ synthetic image.  We then smooth both the synthetic image and Mosaic image to account for differential seeing, and use the biweight of the ratio of these two images to normalize each observation's response curve. 

We construct a single data cube for all HETDEX observations in the area of M101, centered at RA$=210\pdegree 800$ and DEC$=54\pdegree 333$ with spectral pixels of $1\arcsec \times 1\arcsec \times 2$~\AA\null.  To go from the non-uniform sky positions of the fibers to the uniform grid of the data cube, we placed each fiber's flux into the nearest pixel at every wavelength. If multiple fibers from multiple observations contributed to the same pixel, we calculated the median value.  All pixels without a flux contribution from a fiber were masked.  We then performed a Gaussian convolution using a seeing of $1\farcs 8$ to reconstruct the image at each wavelength.  The final data cube is $1201 \times 1201 \times 1036$ pixels.

\subsection{CFHTLS}
\label{sec_cfht}

The Canada-France-Hawaii Telescope Legacy Survey (CFHTLS) is a deep sub-arcsecond (0.8$\arcsec$) wide-field (157 deg$^2$ total) optical survey with {\it u, g, r, i, z} bands. The $7^{th}$ and final release of CFHTLS \citep[CFHTLS-T0007; ][]{2012SPIE.8448E..0MC} is produced by \texttt{TERAPIX} based on a data set collected with MagCam with 36 charge-coupled devices (CCDs) on the CFHT. M101 lies within the footprint of the CFHTLS-T0007. It was observed in {\it i}- and {\it z}-bands on May 11, 2005, and June 7, 2006, respectively. These observations included six consecutive images with 615-second exposures in {\it i}-band and 600-second exposures in {\it z}-band.

\section{Pre-explosion Environments}
\label{sec_envir}

\begin{figure} 
\centering
\includegraphics[width=\textwidth]{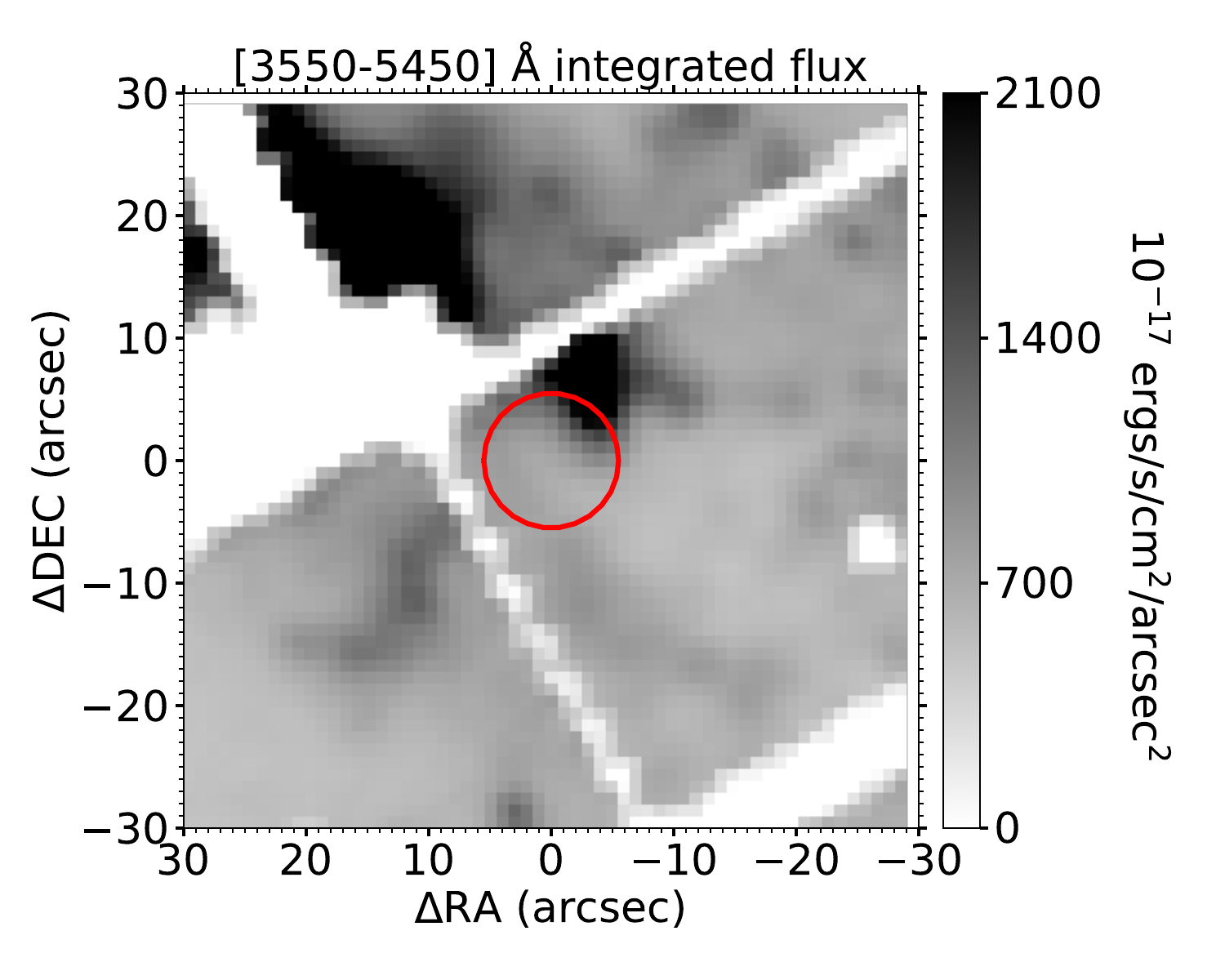}
\caption{The $\pm30\arcsec$ HETDEX image cutout centered on SN~2023ixf taken in the year of 2020. The image is created from the integrated flux within the wavelength range of [3550, 5450] \AA. The pixel size is $1\arcsec\times1\arcsec$. The red circle shows the region of $r<5.5\arcsec$, which is selected for the study of the local environments.}
\label{f_hetdex_g}
\end{figure}

Figure \ref{f_hetdex_g} shows the HETDEX image of SN~2023ixf created from the integrated flux within [3550, 5450] \AA. The bluest wavelengths and the reddest wavelengths are not used to avoid potential edge effects from CCDs. The white gaps are the pixels with no fiber coverage. This $60\arcsec\times60\arcsec$ cutout has a fiber coverage of 80\%. The red circle marks the $r<5.5\arcsec$ region, within which the fiber coverage is 100\%. This region does not include the fibers on the edges of IFUs. The flux of edge fibers can be potentially overestimated due to large aperture corrections. Therefore, the $r<5.5\arcsec$ region is selected for the study of the pre-explosion environments.

\begin{figure*} 
\centering
\includegraphics[width=\textwidth]{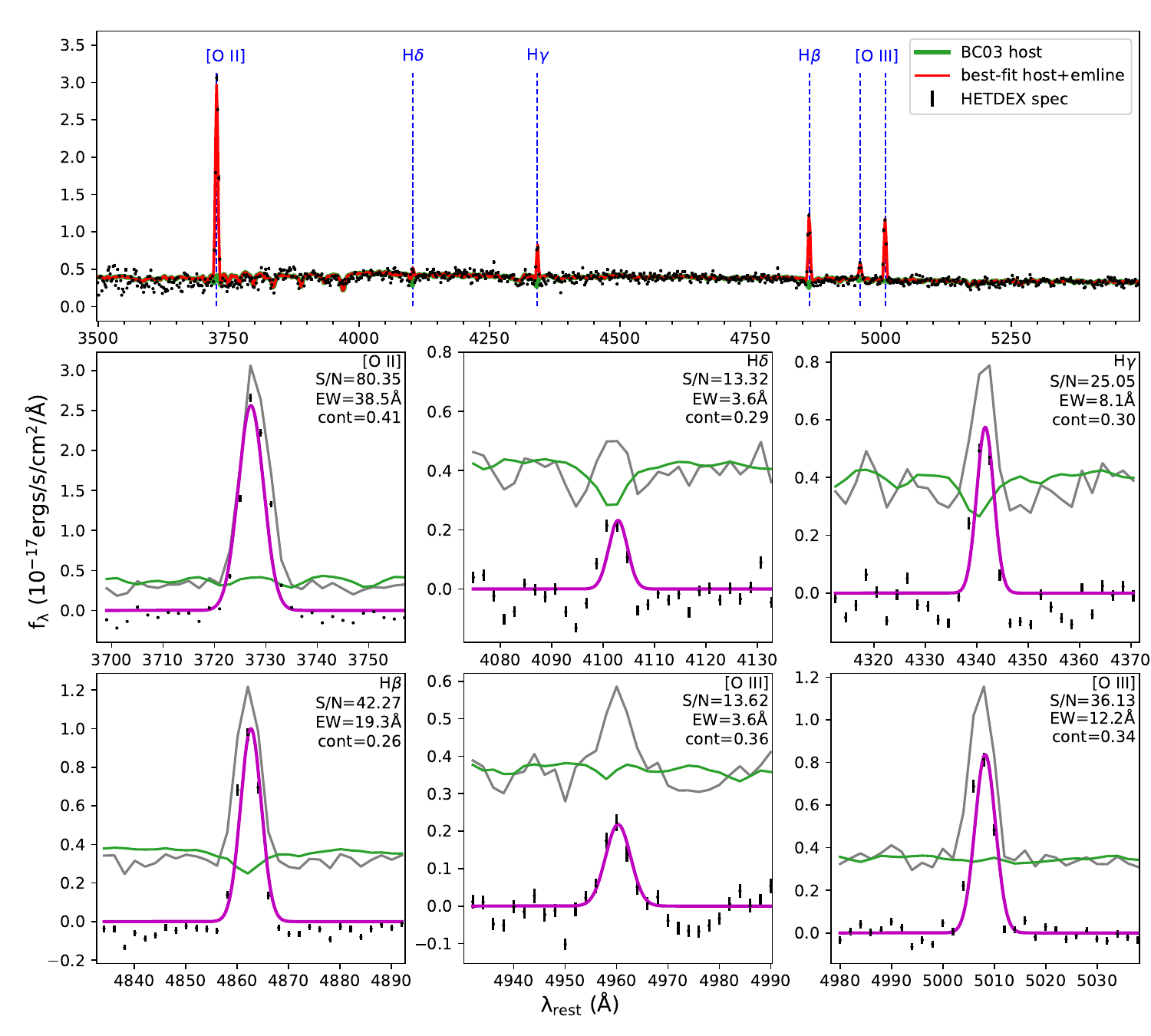} 
\caption{The spectrum of the HETDEX pixel closest to SN~2023ixf. The top panel shows the spectrum in the full HETDEX wavelength range with the black data points. The red line is our best-fit spectrum consisting of the best-fit stellar continuum using BC03 templates and the best-fit emission lines. The blue dashed lines mark the emission lines. The six panels in the 2nd and 3rd rows show the sub-regions of the six emission lines: [\ion{O}{2}] $\lambda3727$, H$\delta$, H$\gamma$, H$\beta$,  [\ion{O}{3}] $\lambda4959$, and [\ion{O}{3}] $\lambda5007$. The green lines are the best-fit stellar continuum in all seven panels. The grey curves in all panels of the bottom two rows are the HETDEX spectrum in each wavelength range. The black data points in the bottom two rows show the continuum subtracted spectra. The magenta lines are the best-fit emission line profiles.}
\label{f_spec_cen}
\end{figure*}

Figure \ref{f_spec_cen} shows the spectrum of the HETDEX pixel closest to SN~2023ixf. The center of this pixel is $0.54\arcsec$ away from SN~2023ixf. As shown by the six panels in the bottom two rows, the nebular emissions of [\ion{O}{2}] $\lambda3727$, H$\gamma$, H$\beta$, [\ion{O}{3}] $\lambda4959$, and [\ion{O}{3}] $\lambda5007$ are significantly detected within our wavelength range. The H$\delta$ emission can be marginally recovered after stellar subtraction using the templates of \cite{BC03}, hereafter BC03. These nebular lines allow the estimates of extinction (Section \ref{sec_extinc}), gas metallicity (Section \ref{sec_metal}), and the star formation rates (Section \ref{sec_sfr}). It is worth noting that no significant variations of the line widths or redshifts/blueshifts of the emission lines are detected as a function of the distance to the SN center with our $\rm 2~\AA$ resolution. \rvs{However, a velocity below our resolution could not be ruled out.}

\begin{figure*} 
\centering
\includegraphics[width=0.45\textwidth]{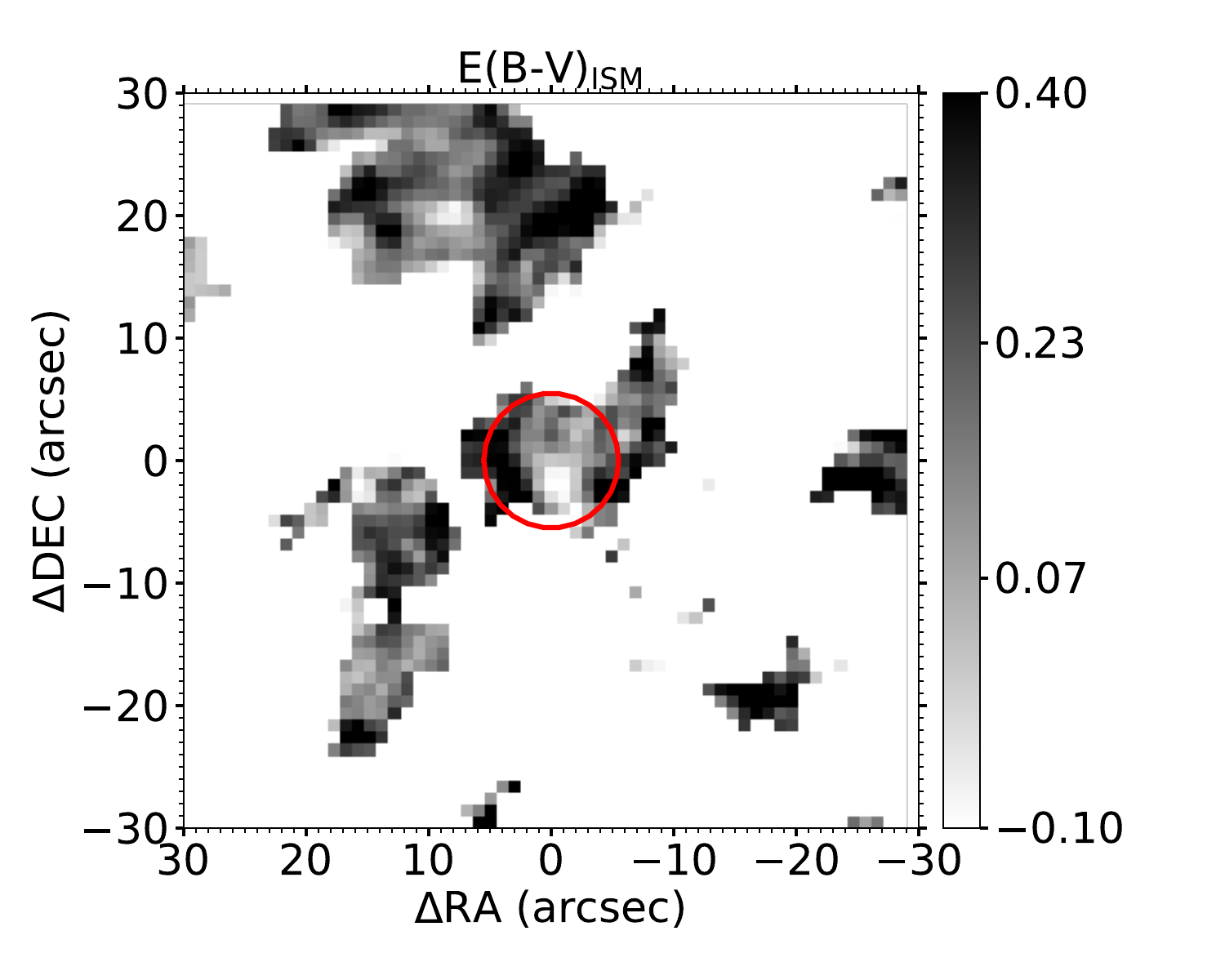}
\includegraphics[width=0.34\textwidth]{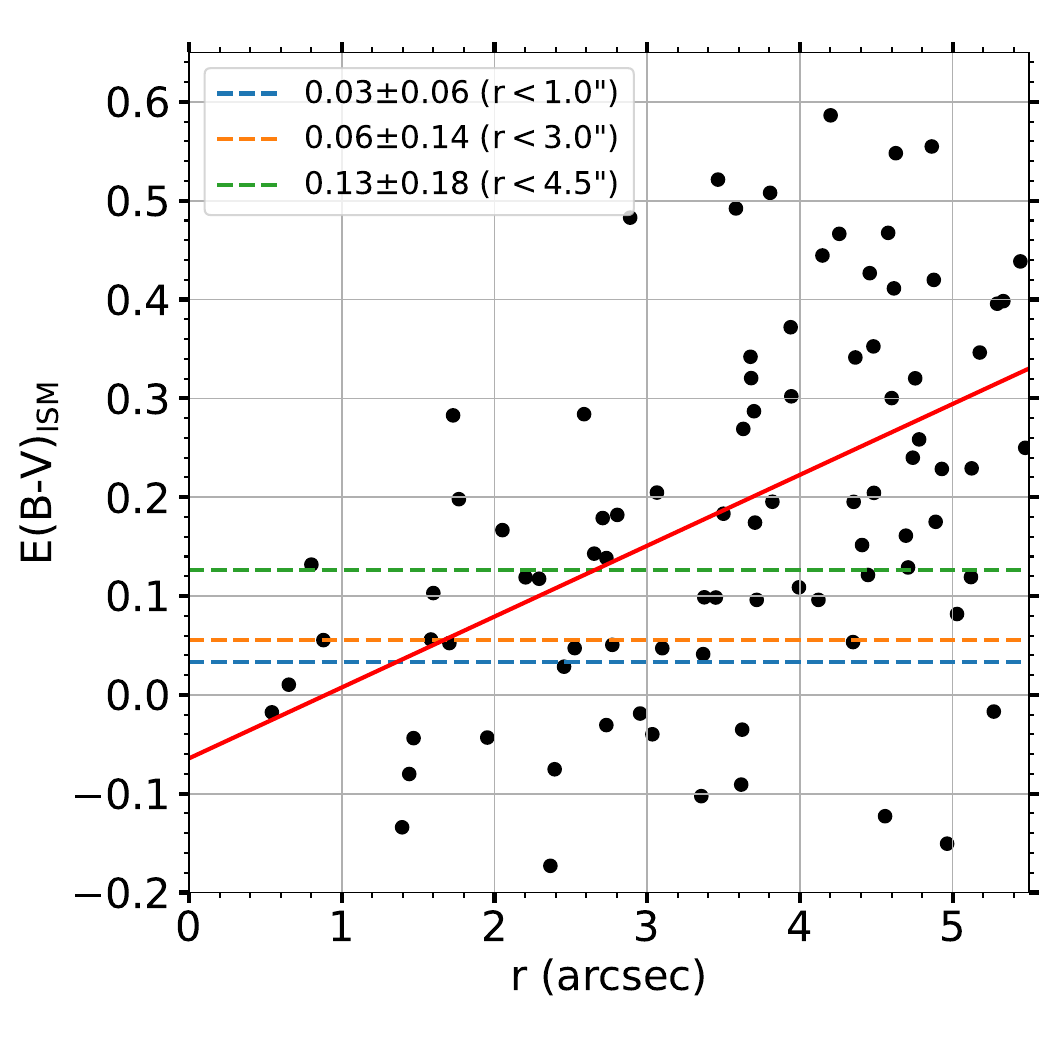}
\includegraphics[width=0.45\textwidth]{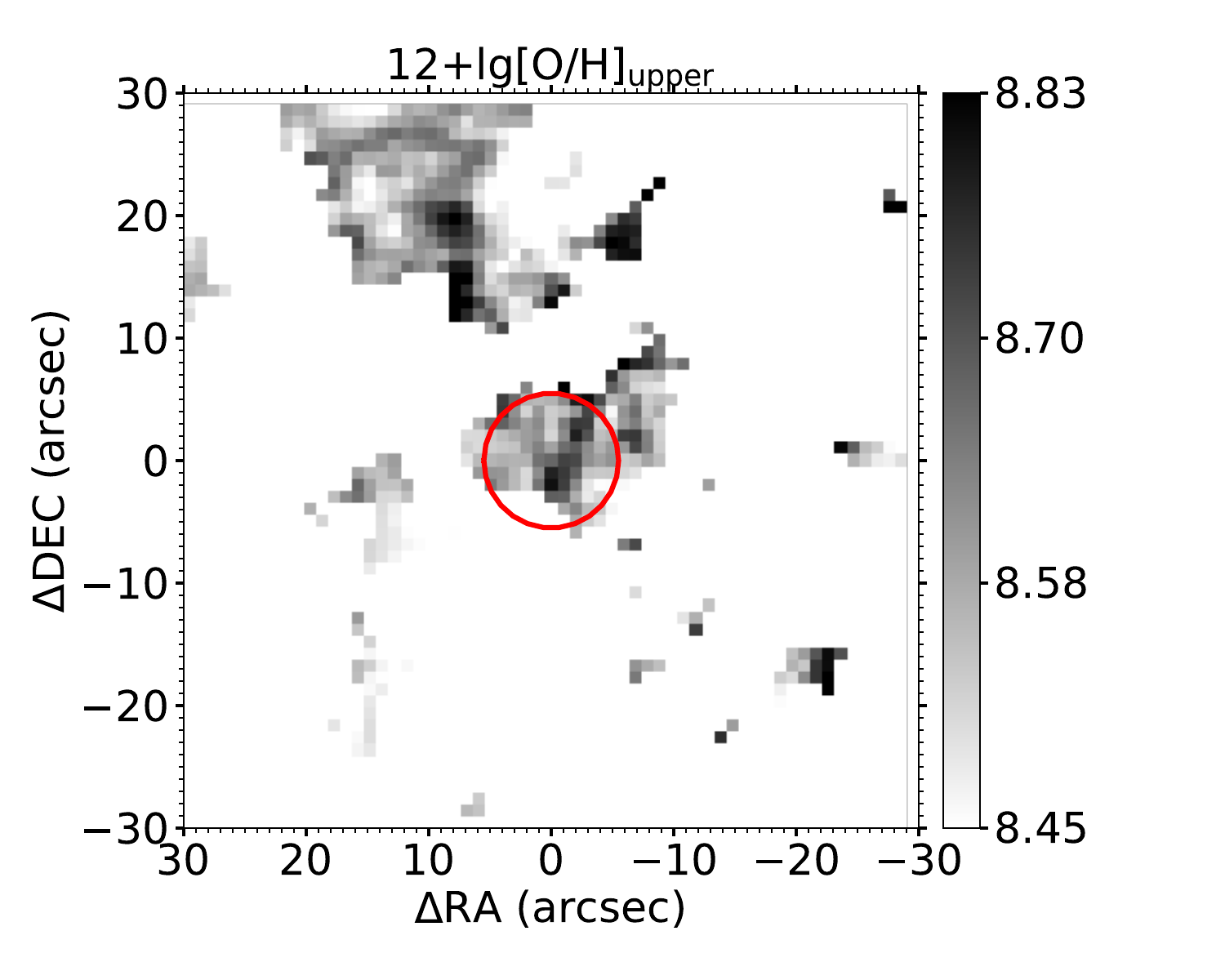}
\includegraphics[width=0.34\textwidth]{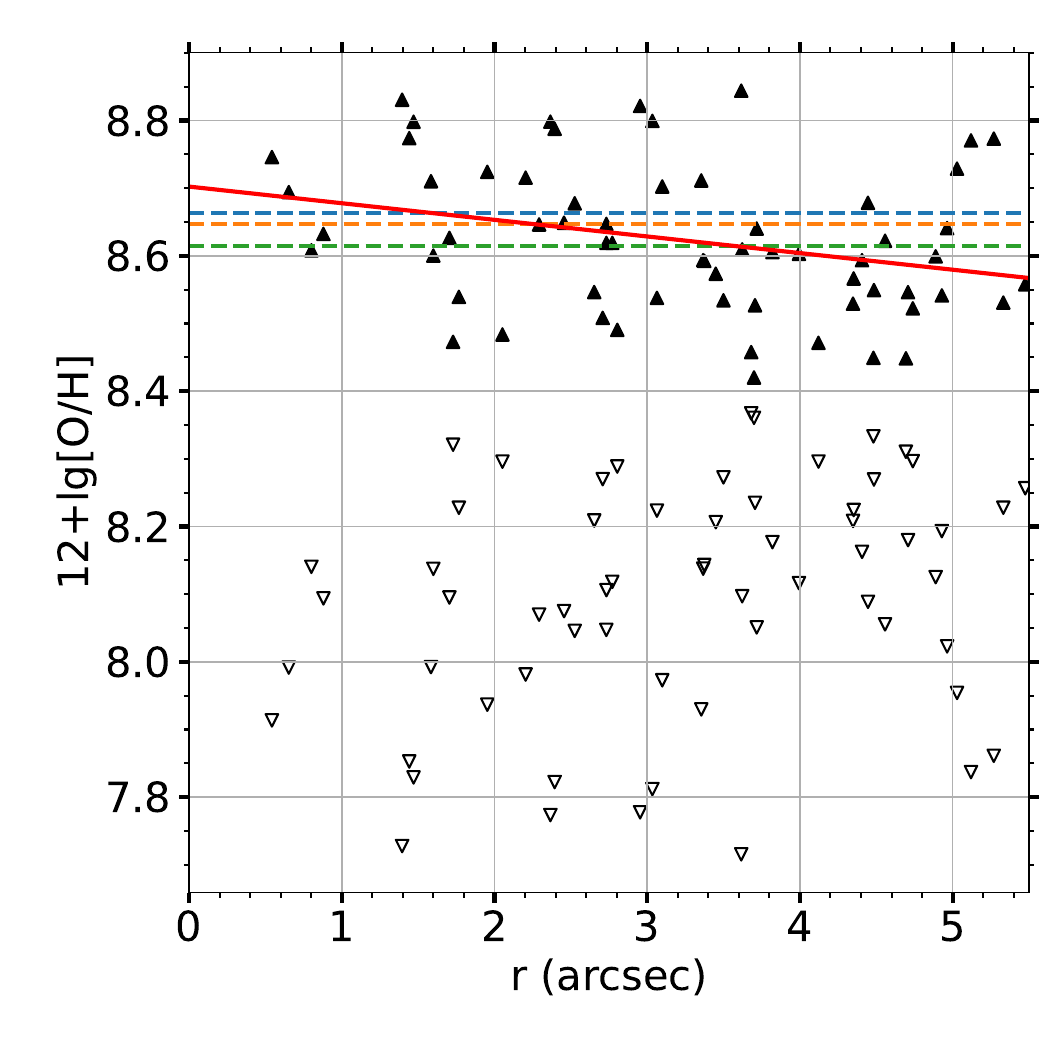}
\includegraphics[width=0.45\textwidth]{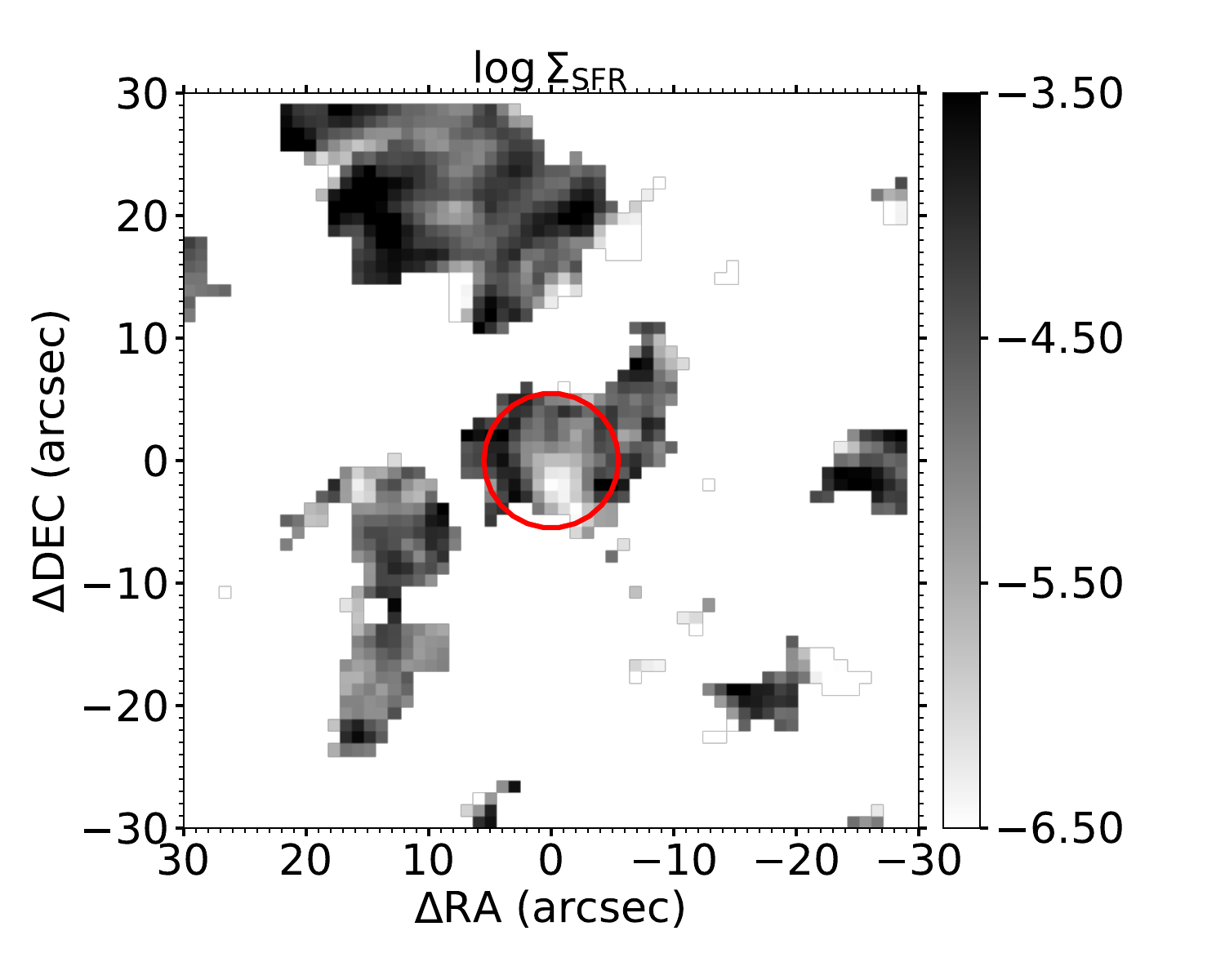}
\includegraphics[width=0.34\textwidth]{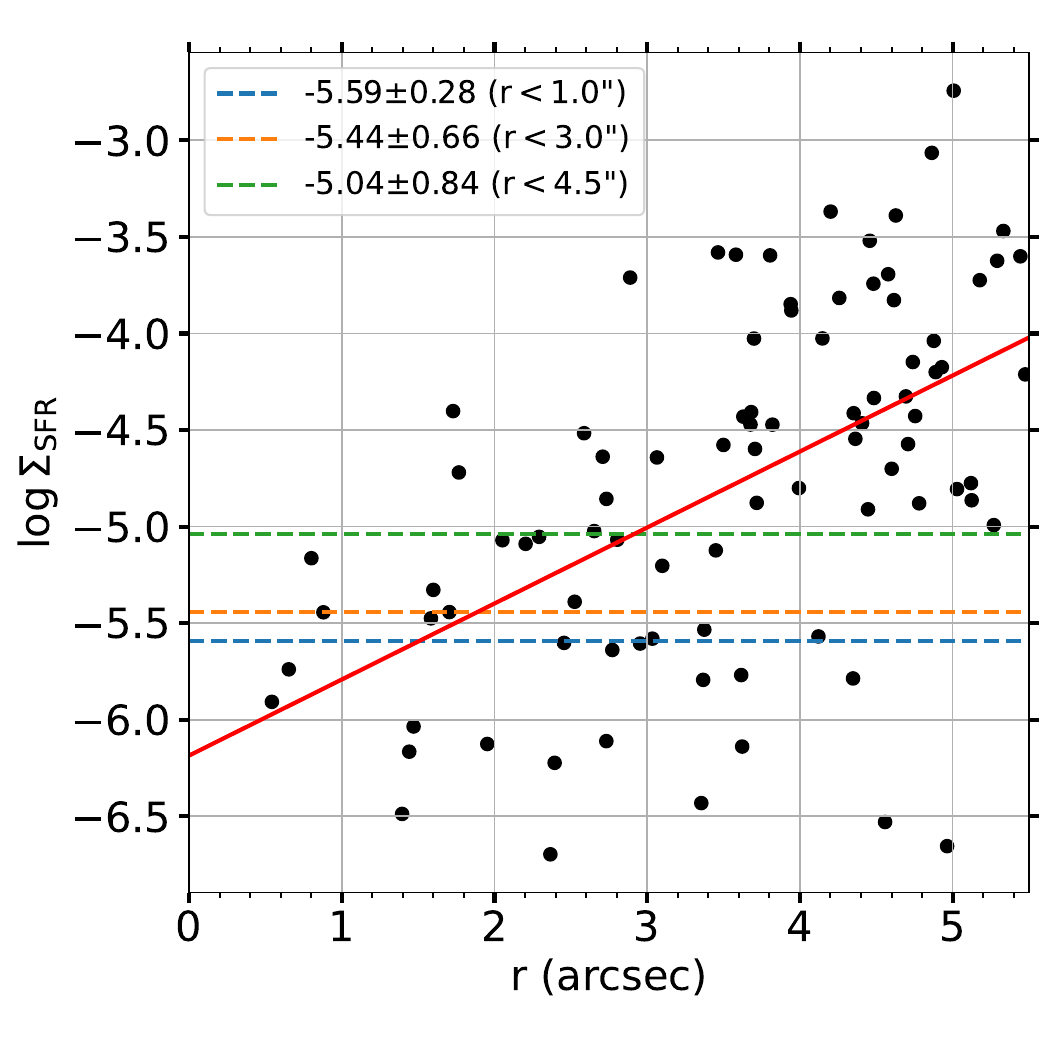}
\caption{The 2-D maps of the diffuse ISM extinction ($E(B-V)_{\rm ISM}$, top left panel), O abundance derived from the metal-rich branch of the $R_{23}$ method (middle left panel), and the surface density of the star formation rate ($\log \Sigma_{\rm SFR}$, bottom left panel) centered on SN~2023ixf. Pixels with fiber coverage but not shown are because their nebular emission lines are not strong enough for reliable measurements of the extinction, O abundance, and the star formation rates, see contexts for details.
Their variations as a function of the separations from SN~2023ixf are shown in the right panels. Data points in the right panels are associated with the individual pixels with measurements within the red circles in the left panels. 
The blue, orange, and green dashed lines in the right panels marks the median values of $E(B-V)_{\rm ISM}$, $\rm 12+log(O/H)_{upper}$, and $\log \Sigma_{\rm SFR}$ of pixels within $r<1\arcsec$, 3\arcsec, and 4.5\arcsec, respectively. The median values are also labeled in the upper left corners. The red solid lines in the right panels are simple linear fits to all pixels within $r<5.5\arcsec$ (the red circles in the left panels) given to guide the eye to radial trends.
In the middle right panel, both the O abundances derived from the the metal-rich branch of the $R_{23}$ method (the solid upper triangles), and those derived from the metal-poor branch (open lower triangles) are shown. 
}
\label{f_envir}
\end{figure*}

\subsection{Extinction}
\label{sec_extinc}

The gas extinction $E(B-V)_{\rm gas}$ can be derived from the Balmer decrement. In this paper, due to the limitation of the wavelength coverage of HETDEX, we use the H$\gamma/$H$\beta$ ratio instead of the more commonly used H$\beta/$H$\alpha$ ratio to calculate the gas extinction. We assume an Galactic extinction curve ($R_V=A_V/E(B-V)=3.1$) \citep{Cardelli1989ApJ} and an intrinsic H$\gamma/$H$\beta$ ratio of 0.47 \citep[$T=15,000~\rm{K}$, case B recombination; ][]{Storey1995}. The reddening suffered by the stellar continuum from the inter-stellar medium (ISM) $E(B-V)_{\rm ISM}$ can then be roughly estimated as 0.44 times the reddening suffered by the ionized gas $E(B-V)_{\rm{gas}}$ \citep{Calzetti2001}. 

Figure \ref{f_envir} shows the extinction map in the upper left panel and its radial variations in the upper right panel. Only pixels with both the H$\gamma$ and H$\beta$ emission lines strong enough, equivalent widths (EWs) greater than $3~\rm\AA$, are calculated with their extinctions and shown. There is a weak increment of the extinction as a function of the separations from SN~2023ixf. The median values of $E(B-V)_{\rm ISM}$ of all pixels within $r<$1\arcsec, 3\arcsec, and 4.5\arcsec\ are $0.03\pm0.06~\rm{mag}$, $0.06\pm0.14~\rm{mag}$, and $0.13\pm0.18~\rm{mag}$, respectively. For M101 exhibiting very high star formation rates, the mean Galactic extinction curve may not be suitable, and the value of $R_V$ can be much larger \citep[e.g. $\sim4.05$;][]{Calzetti1997}. This will give slightly higher measurements of the extinction ($\Delta E(B-V)_{\rm{ISM}}\lesssim0.01~\rm{mag}$), but the weak increasing trend will remain.

Two high-resolution spectra on SN2033ixf was taken with the TK2 cross-dispersed echelle spectrograph (resolving power $R = 60,000$) on the 2.7-m Harlan J. Smith Telescope at McDonald Observatory 2023-05-22.72 and 2023-06-14.66 UT. Resolved components of the narrow NaD doublet at the redshift of M101 ($z=0.000804$) were clearly detected. Equivalent widths of the D1 and D2 components (187.3 and 126.3 m\AA, respectively) correspond to $E(B-V)=0.033\pm0.03~\rm{mag}$ mag using the formulae of \citet{Poznanski12}. This is in perfect agreement with the results of \citet{Lund23, JG23, Smith23} and \citet{Hira23}, and it is not significantly different from our pre-explosion measurement of $E(B-V)=0.03\pm0.06$ in the regions of $r<1\arcsec$. \cite{Niu2023} measures the pre-explosion extinction of the environment within $r<4.5\arcsec$ as $E(B-V)=0.15~\rm{mag}$ using the resolved stars in the HST images. This is consistent with our median $E(B-V)=0.13\pm0.18~\rm{mag}$ in the regions of $r<4.5\arcsec$.

\subsection{Gas Metallicity}
\label{sec_metal}

We use the $R_{23}$ strong-line method to estimate the gas metallicity ($Z$), where $R_{23}$ is a line ratio defined as $\frac{\oii\lambda3727 + \oiii\lambda\lambda4959,5007}{\rm{H}\beta}$\citep[e.g.][]{Tremonti2004}. Line fluxes are dereddened using the gas extinction derived in Section \ref{sec_extinc} and the extinction law of \cite{Cardelli1989ApJ}. For a given $R_{23}$, there are two associated metallicity values, one is in the metal-poor branch, and the other is in the metal-rich branch \citep[e.g.][]{Kobulnicky1999}. SN~2023ixf is $\sim$ 4.5 kpc away from the center of M101, which is within the break radius in the abundance gradient at 15.4 kpc \citep{Garner2022}. This suggests that the surroundings of SN~2023ixf would probably favor the metal-rich branch. However, the possibility of the metal-poor branch can not be fully ruled out. 

Figure \ref{f_envir} shows the metallicity map derived from the metal-rich branch in the middle left panel. In the middle right panel, both the radial variations of the metallicity from the metal-rich branch (solid upper triangles) and those of the metallicity from the metal-poor branch (open lower triangles) are shown. We only calculate the metallicity for pixels that satisfy the following criteria: $\rm EW_{\oii\lambda3727}>3~\AA$, $\rm EW_{\oiii\lambda5007}>3~\AA$, and $\rm EW_{H\beta}>3~\rm\AA$.  Assuming all pixels around SN~2023ixf following the metal-rich branch, \rvs{the median O abundances of all pixels within $r<$1\arcsec, 3\arcsec, and 4.5\arcsec\ are $\rm 12+log(O/H)_{upper} = 8.66\pm0.05$, $8.65\pm0.11$, and $8.61\pm0.11$, respectively. Considering the scattering, the gas metallicitly does not change significantly with their separations from SN~2023ixf. Assuming a solar O abundance of $\rm 12+log(O/H)_{\sun} = 8.69$ and a solar metallicity of $Z_{\sun}=0.013$ \citep{Asplund2021}, the metal-rich branch gives a metallicity of  $Z\sim0.013\pm0.002$ ($\sim Z_{\sun}$), and the metal-poor branch gives $Z\sim0.003\pm0.001$ ($\rm 12+log(O/H)_{lower} = 8.04\pm0.09\ dex$, $\sim 0.25 Z_{\sun}$). We note that considering a more recent solar O abundance estimation of $\rm 12+log(O/H)_{\sun} = 8.77$ and a solar metallicity of $Z_{\sun}=0.0225$ from \cite{Magg2022}, the metal-rich branch would give a metallicity of $Z\sim0.018\pm0.002$, and the metal-poor branch gives $Z\sim0.004\pm0.001$.}

\rvs{\cite{Pledger2023} reported the pre-explosion O abundances of two near the SN site HII regions published in \cite{Kennicutt1996} (region 1098 and 1086 in their Figure 2(b)) as $\rm 12+log(O/H)_{upper} = 8.63$ and $8.59$, which are close to our pre-explosion on-site results observed in the year of 2020. \cite{VanDyk2023} analyzed the after-explosion on-site O abundances to be $\rm 8.43\lesssim12+log(O/H)\lesssim8.86$ from Gemini Spectroscopy on June 3, 2023. The metallicity does not change significantly after the SN~2023ixf event.}

\subsection{Surface Density of Star Formation Rates}
\label{sec_sfr}

The commonly used emission-line indicator of the star formation rate (SFR), the $\rm{H}\alpha$ emission, is out of the wavelength coverage of HETDEX. SFR is therefore estimated using the correlation between the luminosity of [\ion{O}{2}] $\lambda3727$ and SFR \citep[Equation 4 in][]{Kewley2004}. For our case, the calculated SFR of each pixel ($1\arcsec\times1\arcsec$) is the surface density of SFR ($\Sigma_{\rm SFR}$) in units of $\rm M_{\sun}~yr^{-1}~arcsec^2$.

Figure \ref{f_envir} shows the $\Sigma_{\rm SFR}$ map in the bottom left panel and its radial variations in the bottom right panel. Again, only pixels with $\rm EW_{\oii\lambda3727}>3~\AA$ are calculated along with their SFRs. There is an increasing trend of the surface density of SFR as a function of the separations from SN~2023ixf. The median $\log \Sigma_{\rm SFR}$ of all pixels within $r<$1\arcsec, 3\arcsec, and 4.5\arcsec\ are $-5.59\pm0.28$, $-5.44\pm0.66$, and $-5.04\pm0.84$, respectively. 

\section{Mass of the Progenitor}
\label{sec_mass}

\begin{figure*} 
\centering
\includegraphics[width=\textwidth]{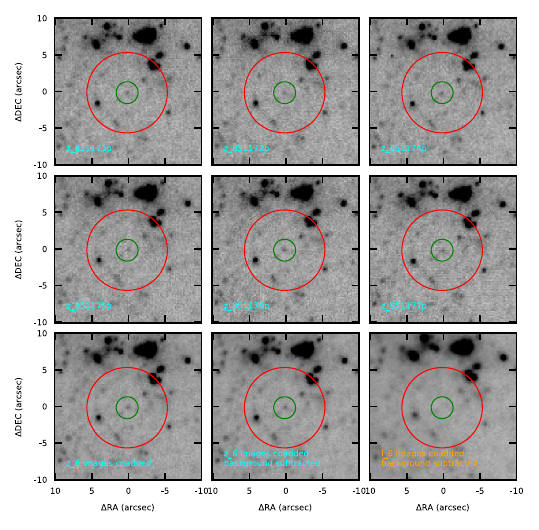}
\caption{The CFHTLS image cutouts centered on SN~2023ixf. The six panels in the top two rows show the six individual $z$-band images. The bottom panels show the $z$-band co-added, the $z$-band background subtracted co-added, and the $i$-band background subtracted co-added images from left to right. The $z$-band observations were taken on June 7, 2006. The $i$-band observations were taken on May 11, 2005. The green and red circles mark the $r<1.5\arcsec$ and $r<5.5\arcsec$ regions of SN~2023ixf, respectively.}
\label{f_cfht}
\end{figure*}

We reprocessed the images of the progenitor from CFHTLS to obtain the magnitudes of the progenitor in the single-exposure images and in the stacked image. Before stacking the images using {\it swarp}\footnote{\url{https://www.astromatic.net/software/swarp/}}, we remove cosmic rays using the python package {\it ccdproc}\footnote{\url{https://github.com/astropy/ccdproc}}. We then recalibrate the astrometry using {\it scamp}\footnote{\url{https://www.astromatic.net/software/scamp/}}, with reference stars from Gaia DR3. {\it SExtractor}\footnote{\url{https://www.astromatic.net/software/sextractor/}} and {\it PSFEx}\footnote{\url{https://www.astromatic.net/software/psfex/}} are used to do the photometry of the progenitor. The photometric zero point is taken directly from the header of the archival fits files. We validate this value using the SDSS survey and the Pan-STARRS survey. 
\rvs{$z_0$ recorded in the header of each individual $z$-band image is 24.754~mag. We cross match each individual image with the SDSS survey and the Pan-STARRS survey. The zeropoint magnitude is then calculated using the common stars recorded in the SDSS survey and the Pan-STARRS survey separately. Take the $z$-band single exposure 851171p as an example: the derived zeropoint from SDSS is $z_0=24.761\pm0.172~\rm{mag}$ and that derived from Pan-STARRS is $z_0=24.753\pm0.176~\rm{mag}$. Both surveys confirm the photometric zeropoint $z_0=24.754~\rm{mag}$ recorded in the header.}
The progenitor is significantly detected in $z$-band at $22.778\pm0.063~\rm{mag}$ in the stacked images (the bottom left panel and the bottom middle panel of Figure \ref{f_cfht}). All six individual $z$-band images show clear detections at the position of SN~2023ixf (top two rows of Figure \ref{f_cfht}). It is worth noting that the progenitor is also significantly detected in the $z$-band observation of the Dark Energy Camera Legacy Survey (DECaLS\footnote{\url{https://www.legacysurvey.org/decamls/}}) at 23.30$\pm$1.58 mag, which confirms our measurements from CFHTLS images.
The progenitor is not detected in $u$-, $g$-, $r$-, and $i$- bands. The $i$-band detection limit is 24.57~mag ($5\sigma$).

Three sources of extinction have been taken into account: the Milky Way \citep[$E(B-V)_{\rm{MW}}=0.008~\rm mag$;][]{Schlegel1998, Schlafly2011}, \rvs{the ISM of} the host galaxy ($E(B-V)_{\rm{ISM}}\sim0.03~\rm mag$; this work), and the circum-stellar medium (CSM) of the progenitor. \rvs{Here we assume $R_V = 3.1$ for all three sources of extinction. We note that $R_V$ can be slightly higher for ISM in the star-forming regions, which will lead to a slightly brighter and redder measurement. Considering the low $E(B-V)_{\rm{ISM}}\sim0.03~\rm mag$, this wouldn't affect the color and magnitude measurements significantly.}
\citet{Niu2023} fits the spectral energy distribution (SED) with the C-rich dust models, and obtains $E(B-V)_{\rm{CSM}}=1.64\pm0.2$\,mag. \rvs{We also note that the C-rich dust model derived extinction should be applied with cautions, as detailed in Section 9.2 of \cite{VanDyk2023}.}
For the CFHTLS $i$- and $z$-bands, $R_i$\rvs{=1.799} and $R_z$=1.299 are adopted \citep{Zhang2023}. \rvs{These together give $A_i\sim3.02~\rm{mag}$ and $A_z\sim2.18~\rm{mag}$.}  The distance modulus of the host galaxy is $29.18$ \citep{2022ApJ...934L...7R}, which gives the absolute magnitudes of the progenitor as $M_i\gtrsim-7.63~\rm{mag}$ ($5\sigma$ detection limit), $M_z=-8.58~\rm{mag}$, and a color limit of $M_i-M_z\gtrsim0.95$.

\begin{figure} 
\centering
\includegraphics[width=8.5cm]{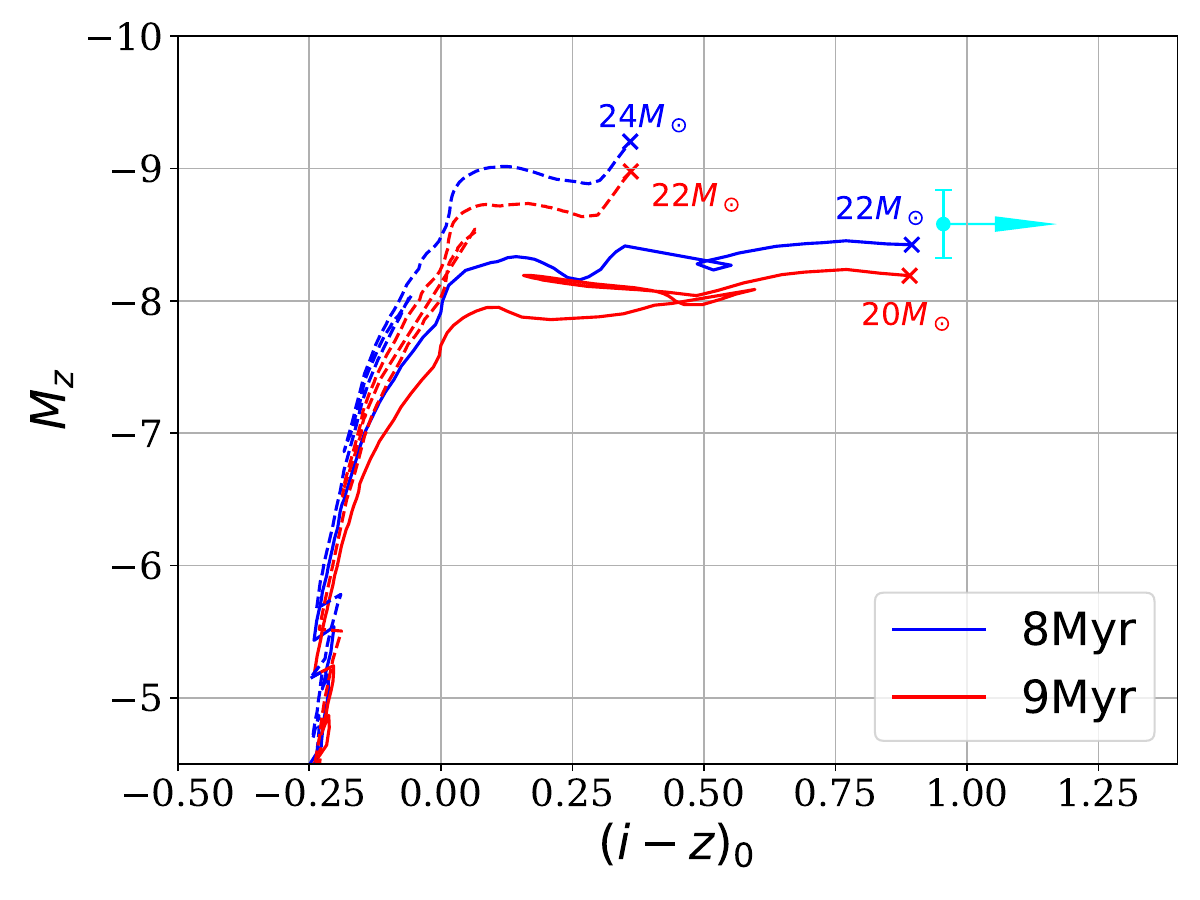}
\caption{Colour-magnitude diagram plotting 8 and 9 Myr \texttt{PARSEC} stellar evolutionary isochrones along with our progenitor candidate. The cyan point is the colour-magnitude position measured from CFHTLS images. The solid lines present the isochrones of solar metallicity $Z_{\sun}$ (the gas metallicity derived from the metal-rich branch using the $R_{23}$ method). The dashed lines present those of $0.25 Z_{\sun}$ (the gas metallicity derived from the metal-poor branch using the $R_{23}$ method).
}
\label{fig:isochrone}
\end{figure}

We use the \texttt{PARSEC}\footnote{\url{http://stev.oapd.inaf.it/cgi-bin/cmd}} stellar evolutionary isochrones \citep{2012MNRAS.427..127B} to estimate the mass of the progenitor (Figure\,\ref{fig:isochrone}). The isochrones that match the CFHTLS colour and magnitude best have an age of \rvs{$\sim$ 8-9~Myr, an initial mass between $\sim20-22~\rm M_\sun$}, and a solar metallicity $\sim Z_{\sun}$.
This metallicity of the progenitor is close to the pre-explosion gas metallicity derived from the metal-rich branch in Section \ref{sec_metal}. We also present the evolution of stars with $0.25 Z_{\sun}$ (the pre-explosion gas metallicity derived from the metal-poor branch in Section \ref{sec_metal}) by the dashed lines \rvs{for comparison reason, although it does not match our color measurement}. \rvs{We note that models with different input evolutionary tracks may result in different initial mass estimates. The widely used \texttt{PARSEC} isochrones do not fit our $(i-z)$ color limit very well. The solar metallicity isochrones stop at $(i-z)_0\sim 0.9$, while our color limit of $(i-z)\gtrsim0.95$ is beyond that. The \texttt{BPASS} \citep{bpass2017} single-star models indicate a lower initial mass range of $\sim15-17~\rm M_\sun$. We also note that our initial mass estimate is close to the maximum possible mass for SN IIP progenitors $\sim20~\rm M_\sun$ \citep{Davies2020}. The $z$-band magnitude combined with an $(i-z)$ color limit may not constrain the evolution track to the best. \cite{Jencson2023} fit the Hubble Space Telescope (HST) photometry and they also concluded with a massive initial mass of $17\pm4~\rm M_\sun$ for SN~2023ixf. \cite{VanDyk2023} analyzed the multi-band data of SN~2023ixf from HST, Spitzer, Herschel, and Wide-Field IR Survey Explorer (WISE), and their SED fitting indicates a less massive initial mass of $\sim12-15~\rm M_\sun$.}

\section{Summary and Discussions}
\label{sec_summary}

In this paper, we study the pre-explosion environments of SN~2023ixf with the HETDEX IFU observations taken in 2020. There are no significant variations of the line widths or redshifts/blueshifts of the emission lines detected as a function of the distance to the SN center with our 2~$\rm \AA$, \rvs{or 133~km/s in velocity space,} resolution. \rvs{We find that SN~2023ixf exploded in a region having low extinction and low SFR locally. This low SFR, however, might be somewhat different from that of the progenitor's birthplace if the progenitor were a runaway star, like e.g. Betelgeuse which has a low velocity of $\sim30$~km/s through ISM.} 
Our pre-explosion measurement of $E(B-V)=0.03\pm0.06~\rm{mag}$ in the regions of $r<1\arcsec$ based on H$\gamma/$H$\beta$ line ratio is not significantly different from our after-explosion measurement of $E(B-V)=0.033\pm0.03~\rm{mag}$ from the resolved NaD doublet. The gas metallicity does not vary significantly with their separations from SN~2023ixf. The metal-rich branch of the $R_{23}$ method gives a near solar gas metallicity $\sim Z_{\sun}$ ($\rm 12+log(O/H) = 8.66\pm0.05~dex$), and the metal-poor branch gives a gas metallicity of $\sim 0.25 Z_{\sun}$ ($\rm 12+log(O/H) = 8.04\pm0.09~dex$). \rvs{If there are future follow-up observations, especially spatially resolved spectroscopic ones, we would learn more about the SN~2023ixf feedback to the host galaxy by comparing with our pre-explosion measurements of the properties of the surrounding ISM.}

We measure the magnitudes of the progenitor of SN~2023ixf in the CFHTLS stack images observed prior to the year of 2010. The progenitor is significantly detected in the $z$-band images at $22.778\pm0.063~\rm{mag}$, but not detected in $u$-, $g$-, $r$-, $i$- bands. A comparatively massive progenitor with an initial mass $\sim22~\rm M_{\sun}$ and a solar metallicity $\sim Z_{\sun}$ is suggested by comparing the extinction-corrected magnitudes with isochrones. If the local gas environments of SN~2023ixf do follow the metal-rich branch of the $R_{23}$ method, this suggests that the metallicity of the progenitor is comparable with that of the surrounding gas.

\vspace{0.4in}
\noindent {\bf Acknowledgments:}
\vspace{0.2in}

The authors thank Xiaoting Fu, Wenyu Xin and Qinghui Sun for valuable suggestions on stellar evolution.

CXL, XLC, HLG, FCK, XKL, JHZ, and XWL acknowledge supports from the ``Science \& Technology Champion Project" (202005AB160002) and from two ``Team Projects" - the ``Innovation Team" (202105AE160021) and the ``Top Team" (202305AT350002), all funded by the ``Yunnan Revitalization Talent Support Program".

HETDEX is led by the University of Texas at Austin McDonald Observatory and Department of Astronomy with participation from the Ludwig-Maximilians-Universit\"at M\"unchen, Max-Planck-Institut f\"ur Extraterrestrische Physik (MPE), Leibniz-Institut f\"ur Astrophysik Potsdam (AIP), Texas A\&M University, The Pennsylvania State University, Institut f\"ur Astrophysik G\"ottingen, The University of Oxford, Max-Planck-Institut f\"ur Astrophysik (MPA), The University of Tokyo, and Missouri University of Science and Technology. In addition to Institutional support, HETDEX is funded by the National Science Foundation (grant AST-0926815), the State of Texas, the US Air Force (AFRL FA9451-04-2-0355), and generous support from private individuals and foundations.

The Hobby-Eberly Telescope (HET) is a joint project of the University of Texas at Austin, the Pennsylvania State University, Ludwig-Maximilians-Universit\"at M\"unchen, and Georg-August-Universit\"at G\"ottingen. The HET is named in honor of its principal benefactors, William P. Hobby and Robert E. Eberly.

VIRUS is a joint project of the University of Texas at Austin,
Leibniz-Institut f\"ur Astrophysik Potsdam (AIP), Texas A\&M University
(TAMU), Max-Planck-Institut f\"ur Extraterrestrische Physik (MPE),
Ludwig-Maximilians-Universit\"at Muenchen, Pennsylvania State
University, Institut f\"ur Astrophysik G\"ottingen, University of Oxford,
and the Max-Planck-Institut f\"ur Astrophysik (MPA). In addition to
Institutional support, VIRUS was partially funded by the National
Science Foundation, the State of Texas, and generous support from
private individuals and foundations.

The authors acknowledge the Texas Advanced Computing Center (TACC) at The University of Texas at Austin for providing high performance computing, visualization, and storage resources that have contributed to the research results reported within this paper. URL: \url{http://www.tacc.utexas.edu}

The research of J.C.W. and J.V. is supported by NSF AST-1813825. J.V. is also supported by
OTKA grant K-142534 of the National Research, Development and Innovation Office, Hungary. 


\end{document}